\documentclass[fleqn,usenatbib,revision]{mnras}

\usepackage{graphicx}	
\usepackage{amsmath}	
\usepackage{amssymb}	
\usepackage[dvipsnames]{xcolor}
\usepackage{hyperref}

\usepackage{newtxtext,newtxmath}
\usepackage{ulem}
\usepackage[T1]{fontenc}

\DeclareRobustCommand{\VAN}[3]{#2}
\let\VANthebibliography\thebibliography
\def\thebibliography{\DeclareRobustCommand{\VAN}[3]{##3}\VANthebibliography}

\begin{document}



\title[On the minimum spin period of accreting pulsars]{On the minimum spin period of accreting pulsars}


\author[\c{C}\i{}k\i{}nto\u{g}lu \& Ek\c{s}i]{
Sercan \c{C}\i{}k\i{}nto\u{g}lu$^{1}$\thanks{E-mail: cikintoglus@itu.edu.tr},
K. Yavuz Ek\c{s}i$^{1}$\thanks{E-mail: eksi@itu.edu.tr}
\\
$^{1}$Istanbul Technical University, Faculty of Science and Letters,
Physics Engineering Department, 34469, Istanbul, Turkey}

\date{Accepted XXX. Received YYY; in original form ZZZ}

\pubyear{2023}

\label{firstpage}
\pagerange{\pageref{firstpage}--\pageref{lastpage}}
\maketitle

\begin{abstract}
The distribution of the spin frequencies of neutron stars in low-mass X-ray binaries exhibits a cut-off at 730 Hz, below the break-up frequency (mass-shedding limit) of neutron stars. The absence of submillisecond pulsars presents a problem, given that these systems are older than the spin-up time-scale. We examine models of disc-magnetosphere interaction near torque equilibrium balanced by the torque due to gravitational wave emission. We note that field lines penetrating the disc beyond the inner radius reduce the maximum rotation frequency of the star, a result well known since the seminal work of Ghosh \& Lamb. We show that the polar cap area corresponds to about half the neutron star surface area at the cut-off frequency if the inner radius is slightly smaller than the corotation radius. We then include the change in the moment of inertia of the star due to the accretion of mass and find that this effect further reduces the maximum rotation frequency of the star. Finally, we include the torque due to gravitational wave emission and calculate its contribution to the torque equilibrium. Our results suggest that all three processes are significant at the cut-off frequency, and that all of them must be considered in addressing the absence of submillisecond pulsars.
\end{abstract}


\section{Introduction}

Accreting neutron stars in low-mass X-ray binary (LMXB) systems rotate very rapidly as indicated by accretion-powered millisecond X-ray pulsars \citep[AMXPs;][]{pat21,dis22} and nuclear-powered burst oscillations \citep{wat12,bha22}.
The spin frequencies of these objects are clustered within the range $\nu \simeq 182-620~{\rm Hz}$ \citep[e.g.][]{pat10,pap+11b,pap+14} and a Bayesian analysis on the spin distribution of known AMXPs yields a cut-off frequency at
\begin{equation}
\nu_{\rm cut-off} = 730~{\rm Hz}
\label{eq:cutoff}
\end{equation}
\citep{cha+03,cha08}.

The neutron stars in these systems are spun up to their high spin frequencies by the transfer of angular momentum via accretion of matter, a process that also reduces their magnetic fields \citep{bis74,bis76}. Millisecond radio pulsars \citep{bac+82} are suggested to descend from these rapidly rotating neutron stars in LMXBs \citep{alp+82,rad82}.
The discovery of accreting millisecond pulsars \citep{wij98}, of radio pulsars with discs \citep{arc+09}, and the existence of transitional millisecond pulsars \citep{pap+13} constitute evidence supporting this `recycling scenario' \citep[see][for reviews]{bha91,dan22}. The fastest rotating radio pulsar discovered to date is J1748-244ad, with a spin frequency of 716 Hz \citep{hes+06}.

Initially, a neutron star rotates rapidly and is spun down by electromagnetic torques, and this is followed by the propeller stage \citep{ill75} where it is spun down by disc torques. The spin-down of a neutron star by such external torques is accompanied by the outward motion of vortex lines in the superfluid component \citep{alp+84}. The magnetic flux tubes are coupled to the vortices, and so they too are carried outwards to the crust where they can be dissipated, leading to a decrease in the magnetic field \citep{sri+90}. It has been argued that this process could reduce the star's magnetic field by three orders of magnitude \citep{jah94}.

As the field decreases sufficiently, the inner radius of the disc approaches the surface of the star, which results in a maximum equilibrium frequency close to the Keplerian frequency at the stellar surface:
\begin{equation}
    \nu_{\rm K}(R) = \frac{1}{2\pi} \sqrt{\frac{GM}{R^3}} = 1973\,\mathrm{Hz} \left( \frac{M}{2\,\mathrm{M}_{\sun}} \right)^{1/2}
    \left(\frac{R}{12~{\rm km}}\right)^{-3/2}~.
\label{eq:nu_K}
\end{equation}
Here, $M$ is the mass and $R$ is the radius of the star.
The break-up rotation frequency, also called the mass-shedding limit, is somewhat lower than the above expression due to the rotational flattening of the star which depends on the internal structure.  
The break-up rotation frequency,  $\nu_{\max}$, of rotating \textit{relativistic} stars is further complicated owing to the relativistic frame-dragging \citep{book_Glendenning97,book_FS13}
through which the rotation of the star drags the space-time with itself.
Many studies have been devoted to determining the maximum rotation frequency \citep{sha+83,sha+89,fri+89,hae89,las96,ste03,lat04,hae+09,don+13,pas17,ria+19,kol20}.
These results show that $\nu_{\max}$ is well above the cut-off frequency given in equation~\eqref{eq:cutoff}. For example, \citet{don+13} provide an analytical function fitting their numerical results (equation~29 in their paper), 
\begin{equation}
\nu_{\max}= 1560~{\rm Hz}\,\left( \frac{M}{2 \mathrm{M}_{\sun}} \right)^{1/2}
    \left(\frac{R}{12~{\rm km}}\right)^{-3/2} - 189~{\rm Hz},
\label{eq:numax}
\end{equation}
which, for a neutron star of mass $M=2\,\mathrm{M}_{\sun}$ and radius $R=12$~km, gives
approximately $1371\,{\rm Hz}$
and we employ this result in the following.
The lack of submillisecond pulsars filling the range $\nu=730-1371~{\rm Hz}$ is a problem, given that the binary lifetime ($10^9~{\rm yr}$) is well above the spin-up time-scale ($10^7-10^8~{\rm yr}$) \citep{whi+88}.

Several arguments have been proposed to explain why the rotation frequencies (and the cut-off frequency) remain below the maximum possible frequency. Before listing these suggestions below, we note that employing $R=12$~km rather than $10$~km reduces this tension by lowering $\nu_{\max}$ from 1862 to 1371~Hz. The recent observational constraints \citep{abb+18,mos+18,mil+19} converge to $R \simeq 12$~km, and 
thus we have used this in scaling equation~\eqref{eq:nu_K} and the following equations in this work.

\citet{li+21} show that the maximum mass accreted by a $1.4\,{\rm M}_{\sun}$ neutron star is about $0.27\,{\rm M}_{\sun}$, while the maximum accreted mass is positively correlated with the initial mass of the neutron star. Furthermore, the mass of the observed second-fastest rotating pulsar is $2.35\pm0.17\,{\rm M}_{\sun}$ \citep{rom+22}. A similar high mass, 
namely $1.9\pm0.3\,{\rm M}_{\sun}$, with a radius of $12.4\pm0.4\,\mathrm{km}$ is estimated for the neutron star in the LMXB 4U 1702-429 \citep{Nattilla17}.
Therefore, we consider the star's mass as $2\,{\rm M}_{\sun}$ in our calculations in this paper.

\begin{figure}
    \centering
    \includegraphics{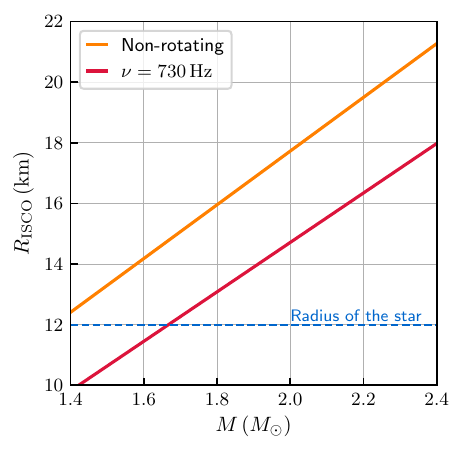}
    \caption{The radius of the innermost circle orbit versus the mass of the star.
    }
     \label{fig:risco}
\end{figure}

In general relativity, there is an innermost stable circular orbit (ISCO) inside which matter cannot rotate on a circular orbit. Therefore, when the inner disc reaches the ISCO, the matter cannot follow the magnetic field lines but will plunge onto the neutron star. Such a system is not expected to show pulsations. For typical neutron stars of $1.4\,{\rm M}_{\sun}$, the ISCO is smaller than the radius of the neutron star and so is not dynamically important for our discussion.
The fastest rotating stars, however, are expected to be massive as a result of mass accretion 
over billions of years.
For massive neutron stars, the ISCO can be greater than the star's radius \citep{luk+18}.
For instance, the radius of the ISCO for a two-solar mass star rotating at $730\,{\rm Hz}$ (see Fig.~\ref{fig:risco}) is
\begin{equation}
    R_{\rm ISCO}\simeq 14.7\,\mathrm{km}
\end{equation}
\citep{mil+98}.

\citet{bil98} argued that losing angular momentum via gravitational wave emission limits the spin frequencies \citep[see][for an updated discussion]{git19}.
Gravitational radiation torques can be strong enough to balance the accretion torque at high frequencies because the gravitational wave radiation increases 
with the fifth power of the angular frequency (see equation~\eqref{eq:torque_gw}).
The quadrupole moments required by gravitational wave emission are caused either by quadrupolar mountains at the star's crust \citep{bil98,ush+00} or by the r-mode instability \citep{and98}.  
On the other hand, accretion torques resulting from thick \citep{and+05} and thin \citep{has11,pat+12} discs are suggested to be sufficient to explain the equilibrium frequencies
\citep[see][for a detailed discussion of solely disc-accretion torque
and additional gravitational radiation torque scenarios based on the spin distributions of neutron stars in LMXBs]{pat+17}.

\citet{bha17} showed that a transient source 
could spin up to higher frequencies than could a persistent source
within the standard disc scenario owing to large deviations from the average mass accretion rate in each outburst. They therefore argued that additional spin-down torques are required to explain the equilibrium spin frequencies.
With the same motive, \citet{dan17} suggested that disc-trapping \citep{dan12} is the significant cause of the spin-down of AMXPs. In this model, the inner radius of the disc is trapped around the corotation radius, where the Keplerian angular velocity of the disc matches with the star's angular velocity, and the star continues to spin down for low accretion rates. As a result, the star ends up with a lower spin frequency after an outburst than in the usual scenarios.

Another possible form of additional spin-down torque is the extraction of rotational kinetic energy by the electromagnetic winds \citep{Par+16}.
According to this model, the stellar magnetic field lines would be opened because of the local angular frequency difference between the disc and the star,
which would be very large in the case of AMXPs. These open magnetic field lines can extract significant rotational energy from the star,
depending on the amount of opened stellar magnetic flux. \citet{Par+16} showed that they could even limit the maximum 
possible frequency of the star to $730\,\mathrm{Hz}$.
Subsequently, these electromagnetic winds were observed in axially symmetric general relativistic magnetohydrodynamic simulations \citep{Par+17,das+22}.

\citet{has+18} investigated whether any unknown state of the matter at high densities might reduce the maximum frequency of the neutron star, finding that the maximum frequency cannot be lower than $\sim 1200\,\mathrm{Hz}$ for any realistic equation of state (EoS). 

Recently, \citet{ert21} showed that the correlation of the final frozen magnetic field of neutron stars with the mass-accretion rate could be responsible for the observed minimum period of the millisecond pulsars.

\citet{book:Tauris23} addressed the problem of the evolution of the binaries. They argued that the high mass accretion rate and low magnetic field of the pulsar do not present in the same stage, because the mass accretion rate is expected to be high in the early time of the binary's life, while the magnetic field of the star decays to low values when the pulsar is old. This suggestion retains the scenario of disc getting close enough to the star to spin up the star above the cut-off frequency.

According to \citet{whi97}, the clustering of the periods indicates that these systems are near torque equilibrium.
While this is a general assumption about these systems, the significance of the critical fastness parameter ($\omega_{\rm c}$ in equation~\eqref{eq:dtorque}) in determining the equilibrium period is not truly appreciated in the literature, and very often, $\omega_{\rm c}$ is set to unity \citep{whi97,has11,pat+12,pat+17,git19}.
The critical value of the fastness parameter essentially determines how much the equilibrium spin frequency can be less than the break-up frequency, 
although it is not the sole effect.

In this paper, we discuss two of the physical processes limiting the equilibrium frequency below $\nu_{\max}$, namely the disc-magnetosphere interaction with some field lines penetrating the disc beyond the corotation radius, and the gravitational radiation torques. Originally,
we show that the changing moment of inertia near torque equilibrium must be taken into account for careful investigation. We assume persistent accretion and thus the accretion rates we quote in this work are representative, average quantities when compared with the works of \citet{bha17} and \citet{dan17}. We first review the basic ideas of the disc-magnetosphere interaction model near torque equilibrium in Section~\ref{sec:Disc_magnetospher_interaction}.
In Section~\ref{sec:dimensionless_torque}, we focus on the critical fastness parameter at which the accretion torque vanishes.
We then consider how the changing moment of inertia would introduce a term that is significant near torque equilibrium
in Section~\ref{sec:momentofinertia}.
We include the torque due to the gravitational radiation in Section~\ref{sec:gwtorque} and investigate how the value of the critical fastness parameter is constrained in this case. Finally, we discuss the implications of our results in Section~\ref{sec:discuss}.

\section{The equilibrium period of accreting pulsars}

We consider the minimum observed (inferred) period is a subject of massive stars, namely $M=2\,\mathrm{M}_{\sun}$.
The radius of the ISCO is $15.4\,\mathrm{km}$ for $M=2\,\mathrm{M}_{\sun}$
and $\nu=730\,\mathrm{Hz}$.
We will consider the $r_{\rm in}\rightarrow R_{\rm ISCO}$ limit as a maximum case for the inner radius of the disc, $r_{\rm in}$.

\begin{figure}
    \centering
    \includegraphics{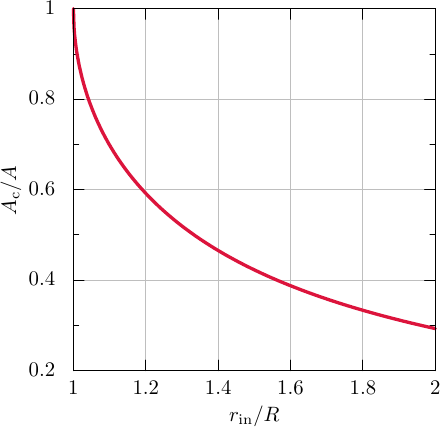}
    \caption{Polar cap area in terms of stellar area versus the inner radius in terms of the star's radius. 
    }
     \label{fig:area}
\end{figure}

To first order, the star's magnetic field is in a dipolar form,
\begin{equation}
     \mathbf{B}=\frac{\mu}{r^3}(2\cos\theta\,\mathbf{\hat{r}}+\sin \theta\,\boldsymbol{\hat{\theta}})\,,
 \end{equation}
 which in polar coordinates can be written as $r=C\sin^2\theta$ by using
\begin{equation}
    \frac{{\rm d}r}{B_{r}}=\frac{r\, {\rm d}\theta }{B_{\theta }}.
\end{equation}
Here, $\mu$ is the magnetic dipole moment of the star at the stellar surface and $C$ labels different field lines, and $C=r_{\rm in}$ for the field line passing through the disc mid-plane ($\theta=\pi/2$) at $r=r_{\rm in}$ \citep[see][p. 26]{book_Ghosh07}:
 \begin{equation}
     r = r_{\rm in} \sin^2 \theta~.
 \end{equation}
This field line intersects the star's surface at $\theta_{\rm c}$, which determines the border of the polar cap: $\sin^2\theta_{\rm c} = R/r_{\rm in}$. The ratio of the area of the two polar caps to the total area $A=4\pi R^2$ is
 \begin{equation}
 \frac{A_{\rm c}}{A} = 1 - \cos \theta_{\rm c} = 1 - \sqrt{1 - \frac{R}{r_{\rm in}}}~.
 \label{eq:area}
 \end{equation}
In Fig.~\ref{fig:area}, we display this ratio as a function of the inner radius in terms of the star's radius. When the area of polar caps is a significant fraction of the total area, the pulsed fraction in the X-ray flux is very small and no pulsations can be observed from the source. In the limit of $r_{\rm in}\rightarrow R_{\rm ISCO}=14.7\,\mathrm{km}$, the $A_{\rm c}/A$ ratio is $0.57$.

\subsection{Disc-magnetosphere interaction}
\label{sec:Disc_magnetospher_interaction}
The X-ray luminosity, $L_{\rm X}$, of the system arises from the accretion of matter onto the neutron star
\begin{equation}
L_{\rm X} = \eta\frac{GM\dot{M}}{R}~,
\label{eq:luminosity}
\end{equation}
where $\dot{M}$ is the mass accretion rate onto the compact object
and $\eta$ is the efficiency coefficient \citep{Ibragimov09}.

We assume that the matter in the disc rotates in Keplerian orbits given by
\begin{equation}
    \Omega_{\rm K} = \sqrt{\frac{GM}{r^3}}~,
\end{equation}
where $r$ is the radial distance.
The critical radius at which the disc matter rotates at the angular velocity of the star, the so-called the corotation radius, is thus defined as
\begin{equation}
    r_{\rm co} = \left( \frac{GM}{\Omega^2}\right)^{1/3}~.
\end{equation}
The inner radius of the disc is where the magnetic stresses are balanced with the material stresses \citep{gho79a,gho79b} and is proportional to the Alfvén radius ($r_{\rm in} =\xi r_{\rm A}$)
\begin{equation}
    r_{\rm in} = \xi \left( \frac{\mu^2}{\sqrt{2GM}\dot{M}}\right)^{2/7}~.
    \label{eq:ralven}
\end{equation}
Here $\xi\sim 1$ is a dimensionless number 
that parametrizes details of modelling the inner parts of the disc
such as the width of the interaction zone of the stellar field,
how the stellar field penetrates into the disc, and the generation of the toroidal field. The magnetic dipole moment of the star at the stellar surface is
\begin{equation}
    \mu = \frac12 B_{\rm d} R^3\, \label{eq:mu}
\end{equation}
where $B_{\rm d}$ is the value of the dipole field at the magnetic poles. 
A dimensionless rotation parameter, the fastness parameter \citep{els77}, is obtained by scaling the angular velocity of the star, $\Omega$, with the Keplerian angular velocity at the inner radius of the disc ($r_{\rm in}$)
\begin{equation}
    \omega_* \equiv \frac{\Omega}{\Omega_{\rm K}(r_{\rm in})}~.
    \label{eq:fastness}
\end{equation}

If the external torque on the neutron star results solely from the interaction of the magnetosphere of the neutron star with the accretion flow, we can write
\begin{equation}
    \frac{\mathrm{d}}{\mathrm{d}t} \left( I \Omega \right) = N_{\rm disc}\,.
\label{eq:torque}    
\end{equation}
The torque acting on a neutron star can be written as
\begin{equation}
    N_{\rm disc} = n \sqrt{G M r_{\rm in}} \dot{M}\,,
 \label{eq:disctorque}   
\end{equation}
where $n= n(\omega_*)$ is the dimensionless torque.

\begin{table}
    \caption{Parameters of the dimensionless torque function for various theoretical models and magnetohydrodynamic (MHD) simulations.
     The fourth column presents the physics limiting the growth of the toroidal field.}
    \begin{center}
    \begin{tabular}{|c|c|c|c|c|}
    \hline
      No. & $\omega_{\rm c}$ & $n_0$ & Note & Ref. \\ \hline
      $1$ & $0.35$ & $1.39$ & Alfv\'en speed & 1 \\
      $2$ &  $0.76$ & $4.4$ & Alfv\'en speed &  2 \\
      $3$ &  $0.71$ & $5.8$ & Turbulent diffusion &  2 \\
      $4$ &  $0.85$ & $8.8$ & Reconnection outside the disc &  2 \\
      $5$ &  $0.76$ & $4.3$ & Buoyancy &  2 \\
      $6$ &  $0.73$ & $6.1$ & Turbulent diffusion &  2 \\
      $7$ &  $0.54$ & $-$   & MHD simulation &  3 \\
        \hline
    \end{tabular}
    \end{center}
     Refs: [1] \citet{gho79a,gho79b}, [2] \citet{Li+96}, [3] \citet{Romanova02}.   
    \label{tab:wcn}
\end{table}

\subsection{The dimensionless torque near equilibrium}
\label{sec:dimensionless_torque}

The torque between two interacting macroscopic systems depends on the velocity difference at the interface. If the magnetosphere interacted only with the inner edge of the disc, one would expect that the accretion torque between the disc and the magnetosphere would depend on $\Omega-\Omega_{\rm K}(r_{\rm in})$, and the equilibrium period is achieved when $r_{\rm in}=r_{\rm co}$ ($\omega_*=1$).
Since the seminal work of \citet{gho79a}, motivated by the existence of systems exhibiting accretion in spin-down, it has been understood that stellar field lines can penetrate the disc in a broad region, and thus the total torque is the integral of contributions interacting with a range of velocity differences. The spin-down torque from the coupling of the stellar field with the disc in the region beyond the inner radius has an important effect on the equilibrium period: the star does not have to spin up until the corotation radius reaches the inner radius. The torque equilibrium is achieved at a somewhat lower value of the fastness parameter, $\omega_{\rm c} \lesssim 1$, called the critical fastness parameter \citep{gho79a,gho79b}. Thus, at equilibrium $r_{\rm in}=\omega_{\rm c}^{2/3}r_{\rm co}$.

Because we are interested only in the near torque equilibrium behaviour, we expand the dimensionless torque into a power series
\begin{equation}
n(\omega_*) = n(\omega_{\rm c}) + n^\prime(\omega_{\rm c})(\omega_*-\omega_{\rm c}) + \frac12 n^{\prime\prime}(\omega_{\rm c})(\omega_*-\omega_{\rm c})^2 \cdots
\end{equation}
where the first term is zero by definition, and the terms higher than the second term can be neglected.
Near the torque equilibrium, the dimensionless torque can thus be written as
\begin{equation}
n = n_0 \left(1 - \frac{\omega_*}{\omega_{\rm c}} \right)
\label{eq:dtorque}
\end{equation}
where $n_0 = -n^\prime(\omega_{\rm c})\omega_{\rm c}$. The values of $\omega_{\rm c}$ and $n_0$ for several models are listed in \autoref{tab:wcn}.

The precise value of the critical fastness parameter depends on the assumptions about the size of the region for which the stellar magnetic fields can penetrate the disc and the physics of how the resulting toroidal field is limited (magnetic diffusion, reconnection, etc.).
While $\omega_{\rm c}=0.35$ in \citet{gho79a}, later work by \citet{Li+96}, with different assumptions on the physics limiting the growth of the toroidal field in the disc, found $\omega_{\rm c} = 0.7-0.85$. A value $\omega_{\rm c} \simeq 0.7$ is inferred by \citet{Tur+17} from observations of quasi-periodic oscillations near torque-reversal for 4U 1626-67 \citep{kau+08}.

Accordingly, we obtain the equilibrium frequency as
\begin{align}
    \nu_{\rm eq} &= \omega_{\rm c} \nu_{\rm K}(r_{\rm in}) \notag \\
         &= \frac{2^{15/14}}{2\pi}\omega_{\rm c} \xi^{-3/2}(GM)^{2/7}R^{-15/7}\eta^{-3/7}L_{\rm X}^{3/7} B_{\rm d}^{-6/7}
         \notag \\
         &= 415\,\mathrm{Hz}\,\omega_{\rm c} \xi^{-3/2} 
        M_{2}^{2/7}R_{12}^{-15/7}\eta^{-3/7}L_{36}^{3/7}B_{8}^{-6/7}
    \label{eq:nueq2BL}
\end{align}
where we referred to equations~\eqref{eq:luminosity} and \eqref{eq:ralven}
and defined $M_{2}=M/2\,\mathrm{M}_{\sun}$, $R_{12}=R/12\,\mathrm{km}$, $L_{36}=L_{\rm X}/10^{36}\,\mathrm{erg/s}$, $B_{8}=B_{\rm d}/10^8\,\mathrm{G}$
because averaged luminosities during outbursts of AMXPs are in the range $10^{36}-10^{38}\,\mathrm{erg/s}$
\citep{pap+11b,pat+12,fal+12,san+17,str+17,san+18a,san+18b,san+22}
and the inferred magnetic fields from long term spin-down evaluations of a few AMXPs are about $10^8\,\mathrm{G}$
\citep{Hartman08,pat09,pat10}.
Equation~\eqref{eq:nueq2BL} is the usual result quoted in the literature (but mostly with $\omega_{\rm c}=1$ pre-set).
The first part of the equation demonstrates the well-known reason the equilibrium frequency of neutron stars in LMXBs is smaller than the Keplerian frequency:
the critical fastness parameter in the numerator $\omega_{\rm c}$ is always less than unity because field lines penetrate the disc beyond the inner radius; that is, the star interacts not only with the inner edge of the disc but also with its slower rotating parts \citep{gho79a,gho79b}.

\begin{figure}
\center
\includegraphics[]{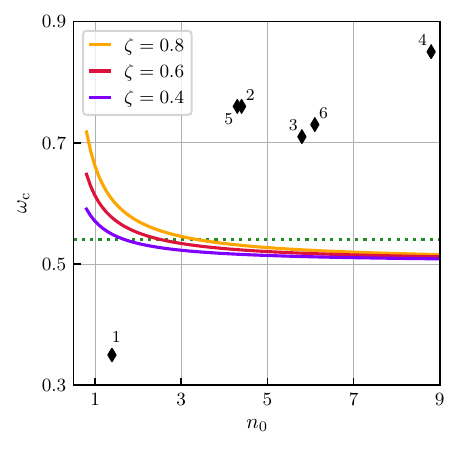}
\caption{Critical fastness parameter, $\omega_{\rm c}$ versus $n_0$ for which $\omega_{\rm c, eff} = \nu_{\rm cut-off}/\nu_{\rm K}(R_{\rm ISCO}) = 0.5$. 
The results are given for a range of $\zeta$ values.
Black diamonds denote the theoretical models
while the green dotted line denotes the result of MHD simulations
in Table~\ref{tab:wcn}.
\label{fig:nw}}
\end{figure}

Considering $r_{\rm in}=R_{\rm ISCO}$, equation~\eqref{eq:cutoff} suggests that $\nu_{\rm cut-off}/\nu_{\rm K}\left(R_{\rm ISCO}\right) \simeq 0.5$ 
for $M=2\,{\rm M}_{\sun}$ and $R=12$~km.
Assuming that the `disc-torque equilibrium model' is the sole cause of the lack of submillisecond pulsars, one would associate $\nu_{\rm cut-off}$ with $\nu_{\rm eq,max}$ and thus write 
\begin{equation}
    \omega_{\rm c} =\nu_{\rm cut-off}/\nu_{\rm K}(R_{\rm ISCO}) \simeq 0.5
\end{equation}
where the numerical value is for $M=2\,{\rm M}_{\sun}$.
Given that the models in the literature bracket this value (see \autoref{tab:wcn}),
the disc-torque equilibrium alone can be considered sufficient to explain the lack of submillisecond pulsars.
As we show in the following subsection, the disc-magnetosphere interaction near torque equilibrium has another contribution arising from the changing moment of inertia of the star. Hence the values of $\omega_{\rm c}$ quoted above are the lowest limit.

\subsection{The effect of the changing moment of inertia of the star}
\label{sec:momentofinertia}
Because accretion can change a star's mass slowly, the star's moment of inertia is expected to change with time. Using
\begin{equation}
    \frac{\mathrm{d}}{\mathrm{d}t} \left( I \Omega \right) = \frac{\mathrm{d}I}{\mathrm{d}t} \Omega + I \frac{\mathrm{d}\Omega}{\mathrm{d}t}\,,
\end{equation}
and
\begin{equation}
    \frac{\mathrm{d}I}{\mathrm{d}t} = \frac{\mathrm{d}I}{\mathrm{d}M} \frac{\mathrm{d}M}{\mathrm{d}t} = \frac{\mathrm{d}I}{\mathrm{d}M} \dot{M}\,,
\end{equation}
equation~\eqref{eq:torque} can be rewritten as
\begin{equation}
    I \frac{\mathrm{d}\Omega}{\mathrm{d}t} = N_{\rm disc} - \frac{\mathrm{d}I}{\mathrm{d}M}\dot{M} \Omega~.
    \label{eq:dotI}
\end{equation}
Although the second term on the right-hand side is well known in the literature and explicit in early work \citep[e.g.][]{lam+73}, it is usually ignored in recent works, possibly because it is small. However, its contribution should become significant when the disc torque is vanishingly small; that is, the system is near torque equilibrium. We thus calculate it and demonstrate its effect on the critical fastness parameter for the first time by employing realistic EoSs. Using equations~\eqref{eq:disctorque} and \eqref{eq:dtorque}, we cast equation~\eqref{eq:dotI} into the form
\begin{equation}
I \frac{d\Omega}{dt} 
= N_0 \left( 1 - \frac{\omega_*}{\omega_{\rm c,eff}} \right)
\end{equation}
where
\begin{equation}
     N_0 = n_0 \sqrt{G M r_{\rm in}} \dot{M}
\end{equation}
is the nominal value of the disc torque
and 
\begin{equation}
\omega_{\rm c,eff} = \omega_{\rm c}  \left(1 + \frac{\zeta \omega_{\rm c}}{n_0}(R/r_{\rm in})^2 \right)^{-1}
\label{eq:w_c_eff}
\end{equation}
is the effective value of the critical fastness parameter. 
Here
\begin{equation}
\zeta \equiv \frac{1}{R^2}\frac{dI}{dM}~,
\label{eq:zeta}
\end{equation}
is a dimensionless value that depends on the EoS of the star.

\citet{bre16} calculated the properties of large numbers of both non-rotating and uniformly rotating 
compact stars in the equilibrium by using 28 EoSs and obtained an empirical formula relating the moment of inertia of the star and its compactness:
\begin{equation}
    I=\left[
      a_1 \mathcal{C}^{-1}
      +a_2 \mathcal{C}^{-2}
      +a_3 \mathcal{C}^{-3}
      +a_4 \mathcal{C}^{-4}
      \right] \frac{G^2M^3}{c^4}~,
      \label{eq:zeta_fit}
\end{equation}
where $\mathcal{C}=GM/Rc^2$, $a_1=0.8134$, $a_2=0.2101$, $a_3=3.175\times10^{-3}$, and $a_4=-2.717\times10^{-4}$.
The largest deviation between the data and their empirical formula is $\sim 0.09$ and its average over all EoSs is $\sim 0.03$.
Their results can be used safely for millisecond pulsars because they employed the Rotating Neutron Stars (\texttt{RNS}) code which solves
full Einstein's equations numerically \citep{Stergioulas95}.
By using this empirical formula,
we calculate $\zeta$ as described in Appendix~\ref{sec:zeta} and find that its value is close to $0.62$ for $M=2~{\rm M}_{\sun}$.

As $r_{\rm in} \rightarrow R_{\rm ISCO}$ we obtain
\begin{equation}
\nu_{\rm eq,\max} = \omega_{\rm c, eff}  \nu_{\rm K}(R_{\rm ISCO})\,, \qquad \omega_{\rm c, eff} =\frac{\omega_{\rm c}}{1 + \beta\zeta \omega_{\rm c}/n_0}~,
\label{eq:nueq}
\end{equation}
where $\beta=R^2/R_{\rm ISCO}^2$.
Note that $\omega_{\rm c,eff}<\omega_{\rm c}$ in all cases.
This result demonstrates a secondary effect causing the lack of sub-millisecond pulsars. The equilibrium spin frequency is achieved at a lower effective value of the critical fastness parameter than what is calculated from the disc-magnetosphere interaction models. This is due to the change in the moment inertia of the star by accretion. Assuming this `modified torque equilibrium model' is the sole cause of the lack of sub-millisecond pulsars, we write 
\begin{equation}
    \omega_{\rm c, eff} =\nu_{\rm cut-off}/\nu_{\rm K}(R_{\rm ISCO})~,
\end{equation}
and thus equation~\eqref{eq:nueq} can be cast as
\begin{equation}
    \omega_{\rm c} =  \frac{\nu_{\rm cut-off}}{\nu_{\rm K}(R_{\rm ISCO})}\left( 1 - \beta\frac{\nu_{\rm cut-off}}{\nu_{\rm K}(R_{\rm ISCO})}\frac{\zeta}{n_0}\right)^{-1}~.
\end{equation}
Theoretical estimations of $n_0$ are between $1$ and $9$ (see Table~\ref{tab:wcn}), 
and the value of $\zeta$ is close to $0.62$ for $M=2~\mathrm{M}_{\sun}$.
In Fig.~\ref{fig:nw}, we show the possible $n_0$ and $\omega_{\rm c}$ combinations for $\nu_{\rm cut-off}/\nu_{\rm K}(R_{\rm ISCO})=0.5$.
Accordingly, the critical fastness parameters given by the theoretical models are greater than the required values to explain 
the cut-off frequency when only the disc torque and the change of the moment of inertia are employed. 
If these theoretical models are valid,
additional spin-down torques such as gravitational radiation are required to limit the maximum frequency of the pulsar.
On the other hand, the critical fastness parameter deduced from the MHD simulations can explain the cut-off frequency 
without the need of additional strong spin-down torques.

\section{The effect of torque due to gravitational radiation}
\label{sec:gwtorque}
\begin{figure*}
    \centering
    \includegraphics{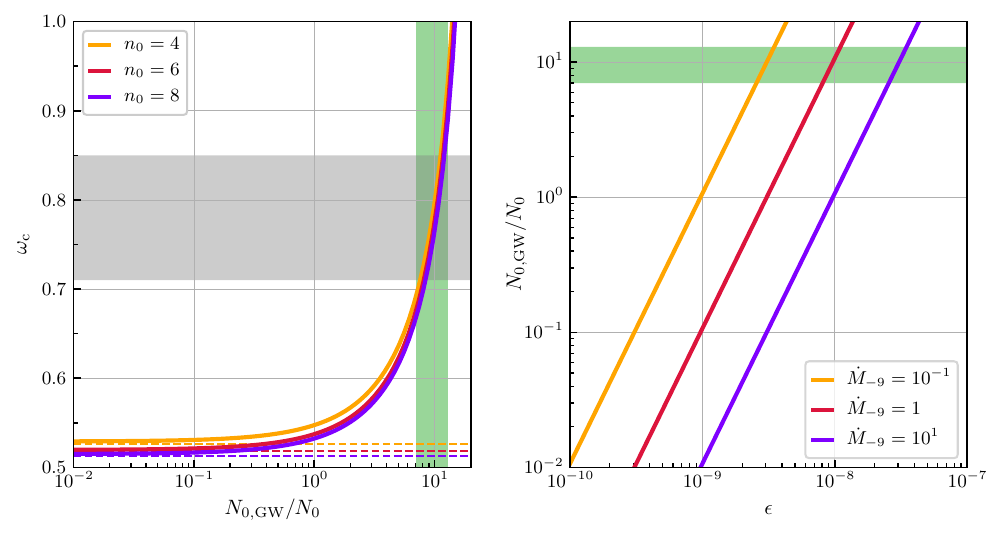}
    \caption{Left-hand panel: The critical fastness parameter versus the $N_{0,\mathrm{GW}}/N_0$ ratio in the presence of a gravitational radiation torque for $M=2\,{\rm M}_{\sun}$.
    In this panel, we assume $\zeta=0.62$. Solid lines correspond to different values of $n_0$.
    Each dashed line denotes the critical fastness parameter of the same coloured solid line in the absence of the gravitational radiation torque (see Fig.~\ref{fig:nw}). The grey region marks the critical fastness parameters 
    of theoretical models enumerated 2-6 in Table~\ref{tab:wcn} while the vertical green
    band marks the corresponding values of $N_{0,\mathrm{GW}}/N_0$.
    Right-hand panel: The ratio of $N_{0,\mathrm{GW}}/N_0$ versus the ellipticity for various values of the mass accretion rate.
    In this panel, we set $n_0=6$ and $M_{2}$, $R_{12}$, $I_{45}$ to one.
    The green band marks where the critical fastness parameter becomes compatible with the theoretical models, as in the left-hand panel.
    }
    \label{fig:alphaw2M}
\end{figure*}

A rapidly rotating neutron star can acquire a quadrupole moment for several reasons ---thermal mountains \citep{bil98,ush+00}, accretion mounds \citep{mel05}, r-mode oscillations \citep{and98,owe+98}--- and emit gravitational waves. As a result of the gravitational wave emission, the star will experience a spin-down torque \citep{book_ST83}
\begin{equation}
    N_{\rm GW}=-\frac{32G}{5c^5}I^2\epsilon^2\Omega^5~,
    \label{eq:torque_gw}
\end{equation}
where $\epsilon$ is the ellipticity of the star.

Accordingly, the spin evolution is determined by 
\begin{equation}
    I \frac{{\rm d}\Omega}{{\rm d}t} = N_0\left(1 - \frac{\omega_*}{\omega_{\rm c,eff}} \right) - N_{0,\mathrm{GW}}\omega_*^5~,
 \label{eq:spin}   
\end{equation}
where
\begin{equation}
    N_{0,\mathrm{GW}} = \frac{32G}{5c^5}I^2\epsilon^2\Omega_{\rm K}^5(r_{\rm in})~.
\end{equation}
The maximum equilibrium frequency, according to equation~\eqref{eq:spin}, is to be found from the solution of
\begin{equation}
    1=\frac{\omega_{\rm eq,max}}{\omega_{\rm c,eff}}+\frac{N_{0,\mathrm{GW}}}{N_0} \omega_{\rm eq,max}^5,
    \label{eq:eq_gw}
\end{equation}
where 
\begin{align}
     \frac{N_{0,\mathrm{GW}}}{N_0} =&\, \frac{32 \epsilon^2}{5 n_0}   \frac{G I^2 \Omega_{\rm K}^4}{c^5 r_{\rm in}^2 \dot{M}}
     =\frac{4.7}{n_0}\epsilon_{-9}^2
      I_{45}^2 \dot{M}_{-9}^{-1} M_{2} ^2
      \left(\frac{r_{\rm in}}{12\,\mathrm{km}}\right)^{-8}\,,
      \label{eq:n0gw}
\end{align}
and $\epsilon_{-9}=\epsilon/10^{-9}$, $I_{45} = I/10^{45}\,\mathrm{g\,cm^2}$, 
and $\dot{M}_{-9}=\dot{M}/10^{-9}\,\mathrm{M_{\sun}\,yr^{-1}}$.
After setting $\omega_{\rm eq,max}=\nu_{\rm cut-off}/\nu_{\rm K}(R_{\rm ISCO})$,
the effective fastness parameter can be determined as
\begin{equation}
    \omega_{\rm c,eff} = \frac{\nu_{\rm cut-off}}{\nu_{\rm K}}
    \left(1-\frac{N_{\mathrm{0,GW}}}{N_0}\frac{\nu_{\rm cut-off}^5}{\nu_{\rm K}^5}
    \right)^{-1}~.
\end{equation}
Hence, the critical fastness parameter in the presence of the gravitational radiation torque
can be written as
\begin{equation}
    \omega_{\rm c} = \omega_{\rm c,eff}
    \left(1-\beta\omega_{\rm c,eff}\frac{\zeta}{n_0}\right)^{-1}~,
\end{equation}
by using the definition of the critical fastness parameter given in
equation~\eqref{eq:w_c_eff} at the $r_{\rm in}\rightarrow R_{\rm ISCO}$ limit.

We calculate $\zeta$ for $M=2\,{\rm M}_{\sun}$ 
by using the empirical formula given in equation~\eqref{eq:zeta_fit} and then
calculate the critical fastness parameter that satisfies the $\nu_{\rm eq, max}=\nu_{\rm cut-off}$
condition, depending on $N_{0,\mathrm{GW}}/N_0$ ratio for various values of $n_0$
in the left-hand panel of Fig.~\ref{fig:alphaw2M}. Accordingly, the required value of the critical fastness parameter
increases as the gravitational radiation torque increases.
The value of the critical fastness parameter becomes consistent with the predictions of the theoretical models
when $7\loa N_{0,\mathrm{GW}}/N_0\loa 13$ for $M=2\,{\rm M}_{\sun}$.

We investigate the possible parameter space for the gravitational radiation torque 
and report it in the right-hand panel of Fig.~\ref{fig:alphaw2M}. 
Only the mass accretion rate and the ellipticity might change a few orders among parameters in the right-hand side of 
equation~\eqref{eq:n0gw}. Therefore, we calculate the ratio of $N_{0,\mathrm{GW}}/N_0$
depending on the ellipticity for various values of the mass accretion rate.

The maximum ellipticity that an accreted crust of a neutron can sustain 
is a long-standing problem. The pioneering calculations of \citet{ush+00} estimated
$\epsilon<10^{-6}$, and later a one order larger upper limit, 
$\epsilon<10^{-5}$, was given by \citet{has+06}.
However, most recent calculations suggest the upper limit of the maximum ellipticity as $\epsilon<10^{-6}$ \citep{Johnson13,Morales22},
with an upper limit even a few orders smaller
estimated by \citet{Gittins21a} and \citet{Gittins21b}.
Furthermore, \citet{Woan18} argue that the spin period derivative of millisecond pulsars
implies a lower limit for the ellipticity, $\epsilon>10^{-8}$. On the other hand,
a tighter upper limit on the ellipticity, namely $\epsilon<10^{-8}$, is obtained
by \citet{abb+20} based on the lack of continuous gravitational wave detection
from five radio pulsars \citep[see also][]{chen20}. 
According to Fig.~\ref{fig:alphaw2M}, the gravitational radiation torque can be effective within the given limits of ellipticity as long as the mass accretion rate is lower than the Eddington limit (i.e.\ $\dot{M} \lesssim 10^{-8}\,{\rm M}_{\sun}\,\mathrm{yr^{-1}}$).

\section{DISCUSSION} 
\label{sec:discuss}

We have studied the equilibrium periods of rapidly spinning low-B accreting pulsars considering both the disc torque and the torque due to gravitational wave emission. 

We first noted that the stellar magnetic field lines penetrating the disc beyond the inner radius reduce the critical fastness parameter, $\omega_{\rm c}$, below unity, a well-known result since the early works by \citet{gho79a,gho79b}. 

We also considered the change of the moment of inertia with mass accretion 
and examined its effect on the spin equilibrium. We have shown that it leads to an effective critical fastness parameter $\omega_{\rm c,eff}$ smaller than $\omega_{\rm c}$. 

This effect shows that spin evolution is possible even when the disc torque vanishes, because accreting matter changes the moment of inertia. Although this is a small effective torque, it becomes significant as the disc torque diminishes near torque equilibrium. 

Finally, we added the gravitational radiation torque and showed that it further limits the maximum frequency that these systems can achieve. The importance of the gravitational wave torque increases with the ratio given in equation~\eqref{eq:n0gw}. Among the parameters in equation~\eqref{eq:n0gw}, the ellipticity and the mass accretion rate are the ones that have a wide range in  parameter space.

Our results constrain the value of the critical fastness parameter to be $0.5$ or slightly higher if the disc torques alone are invoked to address the lack of submillisecond pulsars.
This lower limit is quite close to the critical fastness parameter 
estimated by MHD simulations; however, it is smaller than the results from theoretical models of the disc-magnetosphere interaction.

Because we considered the disc to be near torque equilibrium, the nominal value of the disc torque $N_0= \sqrt{GMr_{\rm in}}\dot{M}$ is multiplied by a small factor $1-\omega_*/\omega_{\rm c,eff}$ which means that $N_0$ needs to attain larger values in order that $N_{\rm disc}$ can balance $N_{\rm GW}$. 

Our results are compatible with the results of \citet{git19} and \citet{pat+17}. These authors, however, use the dimensionless disc torque $(1-\omega_*)$ meaning that they pre-set $\omega_{\rm c}=1$.
That we employ $\omega_{\rm c}<1$ reduces the need for additional spin-down torques such as the gravitational wave radiation torque, hence, 
the required ellipticity is smaller.

The critical fastness parameter is well consistent with the models of the disc-magnetosphere interaction for $7\lessapprox N_{0,\mathrm{GW}}/N_0\lessapprox 13$.
A value of the ellipticity smaller than the observational and theoretical constraints can generate these values of $N_{0,\mathrm{GW}}/N_0$.

Most of the accreting LMXB systems are transients, rather than persistent systems. This implies that the mass accretion rates we use in this work are representative, average quantities. As shown by \citet{bha17} and \citet{dan17}, modelling a transient accretion with 
the instantaneous mass accretion rate provides a higher equilibrium frequency for the star than modelling it with an averaged mass accretion rate. 
Considering this fact, a slightly smaller critical fastness parameter or a stronger gravitational wave torque than those estimated in this work
might be required to explain the absence of submillisecond pulsars.
We plan to investigate this difference by studying the spin and magnetic field evolution of neutron stars under transient accretion
in the future.

\section*{Acknowledgements}
We thank Luciano Rezzolla, Erbil Gügercinoğlu, M.~Ali Alpar and Ünal Ertan for their useful comments. KYE acknowledges support from the Scientific and Technological Research Council of Turkey (TÜBİTAK) with project number 112T105. 
We thank the anonymous referee for constructive comments which helped us to improve the manuscript.
\section*{Data availability}

This is a theoretical paper that does not involve any new data. The
model data presented in this article are all reproducible.

\bibliographystyle{mnras}
\bibliography{refs.bib} 


\appendix

\section{Calculation of $\zeta$}
\label{sec:zeta}

\begin{figure}
\includegraphics[width=\linewidth]{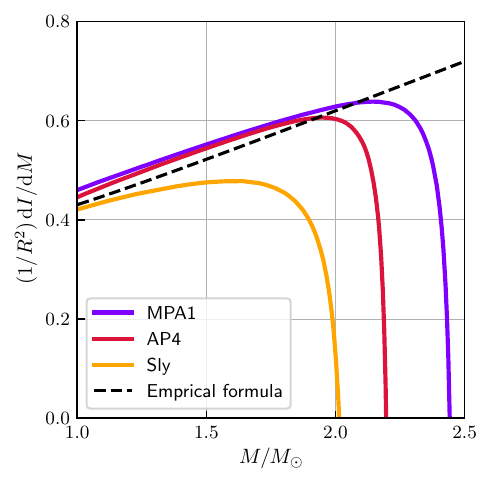}
\caption{Dependence of $\zeta$ on the mass of the neutron star for various EoSs.
The black dashed line represents the result of the calculation 
by the empirical formula given in equation~\eqref{eq:zeta_fit}.
\label{fig:zeta}}
\end{figure}

To determine the value of the dimensionless parameter $\zeta=({\rm d}I/{\rm d}M)/R^2$ given in equation~\eqref{eq:zeta}, 
we use the empirical formula given in \citet{bre16}
(see equation~\eqref{eq:zeta_fit}) and take its derivative with respect to the mass by assuming the radius as a constant. 
Additionally, for comparison,
we have numerically solved the structure of neutron stars for several equations of state (EoSs) as described in \citet{eks+14}. 
In Fig.~\ref{fig:zeta} we report the results of our numerical calculations of $\zeta$ for a stiff EoS \citep[MPA1,][]{ref_MPA1} and two moderate EoSs (\citet[AP4,][]{ref_AP4}; and \citet[SLY,][]{ref_SLy}) together with the estimation from the empirical formula.
We find that $\zeta$ takes values between  $0.4$ and $0.65$ for typical neutron star mass-radius values,
and we estimate $\zeta$ as $0.62$ for $M=2\,{\rm M}_{\sun}$ and $R=12\,\mathrm{km}$.
We neglect the dependence of the radius on the mass in our analytical calculation of $\zeta$ from the empirical formula of \citet{bre16} 
since the radius of the star does not vary too much for a large portion of mass-radius curve.
The estimation of $\zeta$ from the empirical formula departs from the numerical results for the mass values close to the maximum mass supported by the EoS as shown in Fig.~\ref{fig:zeta}, because constant-radius approximation fails when the mass is close to the maximum mass supported by the EoS. Thus, the constant-radius may not be a good approximation for a two-solar-mass star.
If the maximum mass that an EoS compatible with the present mass-radius constraints on neutron stars far exceeds the two-solar-mass, the error due to the constant-radius approximation is small (e.g.\ EoSs MPA1 and AP4). However, if the EoS marginally satisfies the two-solar-mass limit, 
the value of $\zeta$ can be much smaller than the $0.62$ that we employed (e.g.\ EoS Sly).

\bsp	
\label{lastpage}
\end{document}